\documentclass[a4paper,11pt]{article}
\usepackage{graphicx,amssymb,amstext,amsmath}
\usepackage{color}

\definecolor{cbl}{rgb}{0,0,1}                

\newcommand{\bc}{\begin{center}}
\newcommand{\ec}{\end{center}}
\def\ba#1{\begin{array}{#1}\displaystyle}
\newcommand{\ea}{\end{array}}

\newcommand{\beq}{\begin{equation}}
\newcommand{\eeq}{\end{equation}}
\newcommand{\beqa}{\begin{eqnarray}}
\newcommand{\eeqa}{\end{eqnarray}}

\newcommand{\bi}{\begin{itemize}}
\newcommand{\ei}{\end{itemize}}

\def\lt#1{\left#1}
\def\rt#1{\right#1}
\def\t#1{\tilde{#1}}

\def\b#1{\bar{#1}}
\def\frc#1#2{\frac{#1}{#2}}

\newcommand{\Z}{{\mathbb{Z}}}

\newcommand{\C}{{\mathbb{C}}}
\newcommand{\Tr}{{\rm Tr}}

\newcommand{\ep}{\epsilon}
\newcommand{\varep}{\varepsilon}

\def\ba#1{\begin{array}{#1}\displaystyle}

\newcommand{\ket}[1]{\left| #1 \right>} 
\newcommand{\bra}[1]{\left< #1 \right|} 

\begin{document}
\begin{titlepage}
\vspace{0.2cm}
\begin{center}

{\large{\bf{Entanglement Entropy of Non Unitary Conformal Field Theory}}}

\vspace{0.8cm} {\large \text{D.~Bianchini${}^{\bullet}$, O.~Castro-Alvaredo${}^{\bullet}$, B.~Doyon${}^{\circ}$, E.~Levi ${}^{\clubsuit}$ and F.~Ravanini${}^{\spadesuit,\diamondsuit}$}}

\vspace{0.2cm}
{{\small ${}^{\bullet}$} Department of Mathematics, City University London, Northampton Square EC1V 0HB, UK }\\

{{\small ${}^{\circ}$} Department of Mathematics, King's College London, Strand WC2R 2LS, UK}

{{\small ${}^{\clubsuit}$}School of Physics, University Park, Nottingham NG7 2RD, UK}

{{\small ${}^{\spadesuit}$} Dipartimento di Fisica, Universit\`{a} di Bologna, Via Irnerio 46, 40126 Bologna (Italy)}

{{\small ${}^{\diamondsuit}$} I.N.F.N., Sezione di Bologna, Via Irnerio
46, 40126 Bologna (Italy)}

\end{center}

\medskip
\medskip
\medskip

In this letter we show that the R\'enyi entanglement entropy of a region of large size $\ell$ in a one-dimensional critical model whose ground state breaks conformal invariance (such as in those described by non-unitary conformal field theories), behaves as $S_n \sim \frac{c_{\mathrm{eff}}(n+1)}{6n} \log \ell$, where $c_{\mathrm{eff}}=c-24\Delta>0$ is the effective central charge, $c$ (which may be negative) is the central charge of the conformal field theory and $\Delta\neq 0$ is the lowest holomorphic conformal dimension in the theory. We also obtain results for models with boundaries, and with a large but finite correlation length, and we show that if the lowest conformal eigenspace is logarithmic ($L_0 = \Delta I + N$ with $N$ nilpotent), then there is an additional term proportional to $\log(\log \ell)$. These results generalize the well known expressions for unitary models. We provide a general proof, and report on numerical evidence for a non-unitary spin chain and an analytical computation using the corner transfer matrix method for a non-unitary lattice model.
We use a new algebraic technique for studying the branching that arises within the replica approach, and find a new expression for the entanglement entropy in terms of correlation functions of twist fields for non-unitary models.

\medskip
\medskip


\medskip
\medskip
\medskip
\medskip
\medskip
\medskip
\medskip
\medskip
\medskip
\medskip
\medskip
\medskip
\medskip
\medskip

\hfill \today

\end{titlepage}

\section{Entanglement Entropies}
Entanglement is arguably the most non-classical property of a quantum state. It fundamentally determines the performance of classical simulations of many-body quantum systems (e.g.~DMRG \cite{white}) and guides the elaboration of new such simulation algorithms (see e.g.~\cite{vidal}). A physical definition of bi-partite entanglement in quantum systems may be obtained in an operational way. One establishes a preorder: two states' entanglement are compared if the states are related by classical communications, local unitary transformations or entanglement catalysis \cite{bennet,catalysis}. Strikingly, there exists a family of functions that form a faithful representation of this preorder on the set of pure quantum states \cite{majorization}: the R\'enyi entanglement entropies (EEs) $S_n,\;n\geq 1$. Given a state $\ket{\Psi}$ in a Hilbert space ${\cal H}$, and a bi-partition ${\cal H} = {\cal A}\otimes {\cal B}$, the R\'enyi EE $S_n$ is the R\'enyi entropy,
\beq
\label{def2}
    S_n = \frc1{1-n}\log\Tr_{\cal A}(\rho_A^n),
\eeq
of the reduced density $\rho_A = \Tr_{\cal B}\ket{\Psi}\bra{\Psi}$; the von Neumann EE is the special case $S=S_1$. Knowing all the $S_n$ of a state for a bi-partition into subsystems $A$ and $B$ completely determines the entanglement between $A$ and $B$ present in this state.

Besides providing a measure of quantum entanglement, the R\'enyi EEs have a variety of interpretations, which lead to deep connections between quantum entanglement and other physical and geometrical concepts, and are at the source of much activity in this area. For instance, the R\'enyi EEs can be thought of as an entanglement equivalent of the entropy of equilibrium thermodynamics, which gives an order representing the arrow of time. This has physical significance for decoherence, where entanglement with the environment plays a role \cite{lehur}, and in high-energy physics, where the von Neuman EE, which was first introduced within this context \cite{hep}, provides quantum corrections to Hawking's black hole entropy.

Further, it has been observed in recent years that, in the context of many-body systems, the R\'enyi EEs characterize the structure of quantum fluctuations and correlations in a more universal way than other widely studied objects such as correlation functions of local order parameters (see e.g.~\cite{review}). This is especially important near or at critical points, where microscopic interactions give rise to emergent universal collective behaviors described by quantum field theory (QFT), providing some of the most interesting phenomena of theoretical physics.
One-dimensional models are of particular interest in light of modern experimental techniques allowing their realization and precise study \cite{qsc},
as they present surprising anomalous behaviours (e.g. Luttinger liquids, the Kondo effect). One such anomalous effect is seen in the EE.
It was found that the ground state of critical one-dimensional models of infinite length, under a bi-partition where subsystem $A$ is a contiguous set of local degrees of freedom of length $\ell$, has R\'enyi EE that diverges logarithmically at large $\ell$ as \cite{Holzhey,Latorre1,Latorre2,CalabreseCardy}:
\beq
S_n=  \frc{c(n+1)}{6n}\log\left(\ell\right) + O(1)
\label{cft}
\eeq
where the $O(1)$ correction is non-universal. The number $c$ is the ``central charge'', the most important characteristic of the universality class in one dimension, often interpreted as the number of universal local degrees of freedom \cite{cthzam} (e.g. $c=1$ for a Luttinger liquid, $c=1/2$ in the Ising universality class). 
Formula (\ref{cft}) shows an anomalous logarithmic breaking of the area law \cite{arealaw}.
The generalization to system's size $L$ of the order of $\ell$, both large, is \cite{Holzhey,CalabreseCardy}
\beq \label{cftL}
S_n =
\frc{c(n+1)}{6n}\log
\left(\frac{L}{\pi}\sin\lt(\frac{\ell \pi} {L}\rt)\right)+ O(1).
\eeq
Further, off criticality, the EE is finite for $\ell,L=\infty$, but it diverges at large correlation lengths $\xi$ as \cite{CalabreseCardy} $S_n= \frc{c(n+1)}{6n}\log\left(\xi\right) +O(1)$; for the region $A$ extending to the boundary of the system all results are divided by 2; and there are results for excited states \cite{german}. These formulas have been extremely successful: combined with numerical evaluations of the EE using for instance DMRG, they give the most powerful tool for {\em detecting} and {\em distinguishing} critical phases of one-dimensional models, and {\em determining} their universality classes \cite{fab}.

Finally, the R\'enyi EEs also have a striking geometrical interpretation within the duality between universal $d$-dimensional QFT and $d+1$-dimensional statistical field theory that emerges from Feynman's path integral representation. Formula (\ref{def2}), re-interpreted within statistical field theory, gives the R\'enyi EEs as {\em purely geometrical quantities}, which measure the effect of singular geometries on classical thermodynamics \cite{Holzhey}. In particular, for $d=1$, R\'enyi entropies are related to partition functions on spaces with point-like conical singularities at the positions representing the boundary points of the subsystem $A$. That is, singular classical geometries are connected to many-body quantum entanglement. For integer $n$ these conical singularities can in fact be represented using {\em local observables}, the ``branch-point twist fields'' ${\cal T}$ and $\b{\cal T}$ \cite{Knizhnik,CalabreseCardy,entropy}, also defined in quantum chains \cite{permutation}. This re-interprets the R\'enyi EEs as local quantum correlations:
\beq\label{ttbar}
	\Tr_{\cal A}(\rho_A^n) \propto \langle {\cal T}(0)\b{\cal T}(\ell)\rangle.
\eeq
The local-observable representation has been instrumental to the generalization of the above results away from criticality \cite{entropy} and to other entanglement measures \cite{negativity}.

\section{Spontaneous breaking of conformal invariance}

Universal behaviours of one-dimensional critical models with unit dynamical exponent are often described by (1+1-dimensional) conformal field theory (CFT), which possesses space-time conformal symmetry. The results reported above have been obtained within this context. However, a {\em tacit assumption} is that the ground state corresponds to the unique conformally invariant state of the CFT (the ``conformal vacuum''). There are many examples where this assumption fails: {\em conformal invariance may be spontaneously partially broken}. The goal of this letter is to generalize the above to these situations. We provide explicit generalizations of (\ref{cft}) and related formulae, as well as of the local-field representation (\ref{ttbar}).

In order to explain our results and their context, let us recall basic aspects of CFT \cite{cft}. Due to chiral factorization and locality, CFT has an extended algebraic structure based on two independent copies of the Virasoro algebra, with generators $L_n,\,\b L_n,\,n\in \mathbb{Z}$ satisfying the same commutation relations,
\beq
	[L_n,L_m] = (n-m)L_{n+m} + \frc{c}{12}(n^3-n)\delta_{n+m,0}.
\eeq
The number $c$ is the central charge introduced above, which here is interpreted as breaking local conformal invariance. In appropriate units, the Hamiltonian is $L_0+\b L_0 - c/12$. The representation theory of the Virasoro algebra teaches us about the Hilbert space of the quantum system. In particular, one defines the conformal vacuum $\ket{0}$ as the state invariant under all regular conformal transformations, $L_n\ket{0} = 0,\,n\geq -1$, and model-dependent representation theory gives $L_0+\b L_0-c/12$ eigenvalues in correspondence with physical energies.

The physical vacuum may be different from the conformal vacuum, and correspond to a nonzero $L_0$ eigenvalue, when the CFT model is {\em non-rational} (non-rational $L_0$-eigenvalues) and/or {\em non-unitary} (non-unitary representation of the Virasoro algebra).

For instance, certain statistical models, corresponding to non-rational CFT with non-compact ``target spaces'', have been studied in \cite{vjs}, where the conformal vacuum is non-normalizable (like scattering states of quantum mechanics), thus effectively disappearing from the spectrum.
Non-rational CFT is well known to describe various long-standing problems such as two-dimensional quantum particles moving in spatially random potentials \cite{ef,gu} and disordered systems \cite{ms}. For $c=0$, there is a recent study of EE \cite{ali} using a high-energy holography approach. 

Further, critical systems described by non-Hermitian Hamiltonians lead in some cases to non-unitary CFT. Famously, the quantum group invariant integrable XXZ spin chain, with non-hermitian boundary terms, has critical points associated with the minimal models of CFTs with central charges $c=1-\frac{6}{m(m+1)}$ which is negative for $m<2$ rational; see (among others) the study by Pasquier and Saleur \cite{PS}. Another example is provided by the Hamiltonian (with standard Hilbert space structure on $(\C^2)^L$)
\beq
H(\lambda,h)=-\frac{1}{2}\sum_{j=1}^L \left(\sigma_j^z + \lambda \sigma_j^x \sigma_{j+1}^x + i h \sigma_j^x \right), \label{ham}
\eeq
where $\sigma^{x,z}_i$ are the Pauli matrices acting on site $i$ and $\lambda, h \in \mathbb{R}$ are coupling constants.  This model was shown by von Gehlen \cite{vgehlen} to have a critical line in the $\lambda-h$ plane, identified with the Lee-Yang non-unitary minimal model ($c=-22/5$) \cite{yl1}. In these examples, Hamiltonians are non-Hermitian, yet have {\em real energy spectra}. The corresponding real $L_0$ spectra possess {\em negative} eigenvalues: ground states are not conformal vacua.

Non-Hermitian Hamiltonians with real spectra are of particular physical interest and subject to active current research especially in view of the successful application of {\tt PT}-symmetry or pseudo/quasi Hermiticity  \cite{sgh,pt,bender} (for the interplay with integrability see \cite{revpt}). For instance the critical line of the Hamiltonian (\ref{ham}) is related ${\tt PT}$--symmetry breaking \cite{meandreas}. Experimental studies and theoretical descriptions of new physical phenomena have emerged, including optical effects
(unidirectional invisibility, loss-induced transparency) \cite{wow}, transitions from ballistic to diffusive transport \cite{exper}, and dynamical phase transitions \cite{nott,timeint} (in particular using  (\ref{ham})). Interestingly, optical experiments allow {\em experimental access} to non-hermitian quantum mechanics \cite{lon}. Non-Hermitian quantum mechanics is also used in non-equilibrium systems \cite{bw}, quantum Hall transitions \cite{huck}, and quantum annealing \cite{nest}.


\section{Main results and discussion}

Consider a quantum critical chain whose universal behaviour is described by CFT. As mentioned, the energy spectrum follows from the spectra of $L_0$ and $\b L_0$. Assume that their eigenvalues are real, that they have the same lowest eigenvalue $\Delta\neq 0$, and that it is separated from the next higher eigenvalue by a finite amount. That both $L_0$ and $\b L_0$ have the same lowest eigenvalue is the expression that the ground state is translation invariant. Technically, we will also need the lowest-eigenvalue eigenvector of $L_0$ and $\b L_0$ to generate isomorphic Virasoro modules.
The most famous examples of CFT with these properties are the non-unitary minimal series $\mathcal{M}_{p,p'}$ with central charge given before Eq.~(\ref{ham}) with $m^{-1}=p'/p-1$, $p'>p+1$ and $p=2,3,\ldots$; in these cases it is known \cite{cft} that $c_{\rm{eff}}=1-\frac{6}{pp'}>0$. The Hamiltonian (\ref{ham}), on its critical line, corresponds to the Lee-Yang model $\mathcal{M}_{2,5}$.

We look for the R\'enyi EEs of the lowest-energy state of the model. For non-Hermitian Hamiltonians this seems a priori ambiguous, as non-Hermitian operators have the same right- and left-eigenvalues but different eigenvectors. However, we will argue below using {\tt PT} symmetry and chiral factorization of CFT, and see in explicit examples, that at quantum criticality, the eigenvectors corresponding to the common lowest eigenvalue of $H$ and $H^\dag$ {\em are equal to each other}. With this unique vector and the standard Hilbert space structure, the R\'enyi EE is well defined and unambiguous. {\em Our first main finding is that (\ref{cft}) and related formulae hold with the replacement $c\mapsto c_{\rm eff}:=c - 24\Delta>0$}. We note that formula (\ref{cft}) would not make physical sense when $c<0$, such as in the Lee-Yang model corresponding to (\ref{ham}). This is reminiscent of the work of Itzykson, Saleur and Zuber \cite{ceff} who showed that the {\em effective central charge} $c_{\rm eff}$ replaces $c$ in the expression of the ground state free energy found by Affleck \cite{affleck} and Bl\"ote, Cardy and Nightingale \cite{BCN}. Our second  finding is that there is an additional term proportional to $\log(\log \ell)$ if the lowest $L_0$-eigenspace  is ``logarithmic''; that is, if $L_0$ has, on this space, the form $\Delta I + N$ where $I$ is the identity matrix and $N$ is nilpotent of some degree $r$.

The finding that the effective central charge is involved instead of the central charge has important consequences: (i) the values of $c_{\rm eff}$ (not $c$) provide an order amongst critical ground states in agreement with the ordering of their entanglement; and (ii) the determination of a critical phase by the numerical evaluation of the EE gives access to $c_{\rm eff}$ (not $c$). These consequences further point to the fact that in general $c_{\rm eff}$, and not $c$, counts the physical, universal number of local degrees of freedom. These findings also open a way to the elusive proof of RG monotonicity of $c_{\rm eff}$, based on entanglement concepts, generalizing the proof in unitary models \cite{cthzam} based on unitary QFT.

The fact the $c_{\rm eff}$ occurs in place of $c$ in important physical quantities suggests that a theory for local physical observables should involve $c_{\rm eff}$ instead of $c$ more prominently. Unfortunately, such a theory does not exist yet. Nevertheless, the {\em classical singular geometry} associated to R\'enyi EEs gives us indications. According to the state-field correspondence of CFT, every state is associated to a local field. Let $\phi$ represent the local field associated to the lowest-energy state. Appealing to the operator product expansion, one may form, from the branch-point twist field ${\cal T}$, the {\em composite field}  \cite{ctheorem}
\beq\label{tphi}
	:\!\!\mathcal{T}\phi\!\!:(x):=n^{2\Delta-1} \,\lim_{\varepsilon \rightarrow 0} \varep^{2(1-\frc1n)\Delta} \mathcal{T}(x+\varep) \phi(x).
\eeq
Here the factor $n^{2\Delta}$ ensures that the field has the standard CFT normalization. Using the above findings, we will show that the generalization of (\ref{ttbar}), representing conical singularities and the R\'enyi entropies for integer $n$ using local fields, is
\beq
\Tr_{\cal A}(\rho_A^n) \propto  \frac{\langle:\!\!\mathcal{T}\phi\!\!:(\ell) :\!\!\b{\mathcal{T}}\phi\!\!:(0)\rangle}{\langle \phi(\ell)\phi(0)\rangle^n}. \label{newt}
\eeq
This and related formulae have important potential applications to the EE off criticality \cite{inprogress} and the entanglement negativity \cite{negativity}, and suggests a path towards defining physical correlation functions in non-unitary models.

\section{ CFT derivation}
In this section we use new techniques related to, but simplifying, those of \cite{Holzhey}. For simplicity we consider $\Tr_{\cal A}(\rho_A^n)$ in the ground state of a critical chain with a boundary, the region $A$ extending a distance $\ell$ from the boundary. Within the replica trick, one write $\Tr_{\cal A}(\rho_A^n) = Z_n/Z_1^n$, where $Z_n$ is the partition function of a Euclidean CFT composed of $n$ copies of the original model, connected to each other cyclically along a cut representing the region $A$ \cite{Holzhey}. Here the path integral lies on the half-plane $\{z:{\rm Re}(z)>0\}$ and the cut runs from the origin to the point $\ell$. Around the $2\pi n$-angle conical singularity at $z=\ell$, the CFT description breaks down and the lattice structure becomes important, thus
we introduce an ultraviolet cut-off $0<\ep\ll \ell$. After the conformal transformation $z\mapsto w = i \log\lt(\frc{\ell-z}{\ell+z}\rt)$ to the half-cylinder, the cut runs from ${\rm Im}(w)=\log\ep/\ell$ (for $\epsilon \rightarrow 0$) to ${\rm Im}(w)=0$. Quantizing with time running along the cut, we have
\beq\label{ab}
	Z_n=\bra{a}e^{-\log(\frac{\ell}{\ep})\,H_{\rm orb}} \ket{b}, \, \,
	Z_1^n=\bra{a}e^{-\log(\frac{\ell}{\ep})\,H_{{\rm rep}}} \ket{b}
\eeq
where $H_{\rm orb}$ is the ``orbifold'', and $H_{\rm rep}$ is the ``replica'' Hamiltonian; $\bra{a}$ and $\ket{b}$ are conformal boundary states, the precise form of which is not important. The difference between $H_{\rm orb}$ and $H_{\rm rep}$ is in the continuity condition of the energy density: the chiral stress tensor $T^{(j)}(w)$ on the $j^{\rm th}$ copy is, respectively, cyclic, $T^{(j)}(w+2\pi) = T^{(j+1)}(w)$, or periodic, $T^{(j)}(w+2\pi) = T^{(j)}(w)$. In both cases the total stress tensor $T(w)=\sum_{j=1}^n T^{(j)}(w)$ gives rise to the Hamiltonian.

Fourier modes of stress tensors generate Virasoro algebras. With periodic conditions there are $n$ commuting Virasoro algebras $L_k^{(j)}$, $k\in\Z$ each of central charge $c$. The total Virasoro algebra is
$
	L_k^{{\rm rep}}=\sum_{j=1}^n L_k^{(j)}
$ and $H_{{\rm rep}} = L_0^{{\rm rep}} + \b L_0^{{\rm rep}} -nc/12$. On the other hand, in the cyclic case, there is a field that is continuously winding around the copies, $T_{\rm orb}(w) = T^{(1+[w/2\pi])}(w\, {\rm mod}\, 2\pi)$, generating a single Virasoro algebra ${\mathcal{L}}_{k}$, $k\in\Z$ of central charge $c$. The total Virasoro algebra is then a subalgebra of central charge $nc$ generated by ${\cal L}_{nk}$, $k\in\Z$. Doing the Fourier transform one finds 
\begin{equation}\label{LL}
	L_k^{{\rm orb}} = \frac{\mathcal{L}_{n k}}{n}+\Delta_{\mathcal{T}} \delta_{0,k},
\eeq
and we have $H_{{\rm orb}} = L_0^{{\rm orb}} + \b L_0^{{\rm orb}} -nc/12$. Equation (\ref{LL}) has been found in the context of the study of orbifold CFT, see e.g.~\cite{KacWakimoto,Bouwknegt,Borisov} and, more recently, in connection with the investigation of the energy flow in critical systems out-of-equilibrium \cite{DDH}. On the right-hand side of (\ref{LL}) the shift is the dimension $$\Delta_{\mathcal{T}}=\Delta_{\t{\mathcal{T}}}=\frc{c}{24}\lt(n-\frc1n\rt)$$ of the branch-point twist fields \cite{Knizhnik,CalabreseCardy} .

Since the lowest $L_0^{(j)}$ and ${\cal L}_0$ eigenvalues are $\Delta$, in the limit $\ell/\ep\to\infty$ we find $Z_n\propto e^{-2\log(\ell/\ep)\,(\tilde{\Delta}-nc/12)}$ and $Z_1^n\propto
e^{-2\log(\ell/\ep)\,(n\Delta-nc/12)}$ where
\beq\label{dtp}
	\tilde{\Delta}= \frac{\Delta}{n} + \Delta_{\mathcal{T}}
\eeq
Therefore, $\Tr_{\cal A}(\rho_A^n) \sim \lt(\frc\ep \ell\rt)^{
    \frc{c_{\rm eff}}{12} \lt(n-\frc1n\rt)}$
whereby we obtain
\begin{equation}
S_n=\frac{c_{\rm{eff}}(n+1)}{12 n}\log \frac{\ell}{\epsilon} \label{nus}
\end{equation}
up to finite non-universal additive terms.
Similar calculations can be performed for $A$ a region of length $\ell$ in an infinite chain, for system lengths $L\propto \ell$, and for off-critical systems at large correlation length, reproducing (\ref{cft}) and related formulae with the replacement $c\mapsto c_{\rm eff}$, as claimed.

Notice that the composite field $:\!\!\mathcal{T}\phi\!\!:(x)$ (\ref{tphi}) has conformal dimension (\ref{dtp}) \cite{ctheorem}. Hence from pure scaling dimension analysis the result above (\ref{nus}) can be recast into $\Tr_{\cal A}(\rho_A^n) \propto  \frac{\langle:\mathcal{T}\phi:(\ell)\rangle}{\langle \phi(\ell)\rangle^n}$, and for $A$ a region of length $\ell$ in an infinite chain, to  (\ref{newt}). This provides the local-field representation of the R\'enyi entropy and of conical singularity in non-unitary models, as claimed. Indeed, for non-unitary theories $:\!\!\mathcal{T}\phi\!\!:$ is the lowest-dimension field with the correct twist property, hence represents the integer-angle conical singularity. We emphasize that the elegant techniques used in \cite{CalabreseCardy}, where fields associated to conical singularities are studied using conformal transformations, {\em fail} in the non-unitary case; a full understanding of such techniques is still missing.

The above derivation is easily adaptable to the cases where $L_0$ (here standing for ${\cal L}_0$ or $L_0^{(j)}$ for any $j$) is of ``triangular form'' on its lowest eigenspace, corresponding to a logarithmic representation of the Virasoro algebra. Assume $L_0$ takes, on its lowest eigenspace, the form  $\Delta I + N$ where $I$ is the identity matrix and $N$ is nilpotent of degree $r$: $N^{r}=0$. Let $0<p\leq r-1$ be the largest integer power such that $\langle a | N^{p}| b \rangle\neq 0$. Then $e^{-u L_0} = e^{-u\Delta I}\lt(\sum_{k=0}^{r-1} (-uN)^k/k!\rt)$. With $u=\log(\ell/\ep)\to\infty$ evaluating $\langle a| e^{-uL_0)}|b\rangle$ and keeping only the leading power $\propto u^{p}$ in the sum we find
\beq\label{reslog}
S_n=\frac{c_{\rm{eff}}(n+1)}{12 n}\log
\frac{\ell}{\epsilon} + p
\log\lt(\log\frc{\ell}{\ep}\rt)
\eeq
again up to finite non-universal additive terms. Again, we expect that this could be understood in terms of the field $:\!\!\mathcal{T}\phi\!\!:(x)$, which is now logarithmic, but this is beyond the scope of this paper. The entanglement entropy of logarithmic CFT has been considered in \cite{ali} from a holographic point of view, although the situation there seems sligthly different from that considered here.

Finally we, make a technical note on an aspect of the derivation which requires clarifications from the viewpoint of non-Hermitian quantum mechanics. We observe that Euclidean QFT, as used in our derivation, naturally associates the lowest-energy right- and left-eigenstates $\ket{\Psi_R}$ and $\bra{\Psi_L}$ to negative and positive infinite times, respectively -- hence the replica trick, a priori, actually evaluates the incorrect quantity $\Tr_{\cal A}(\t\rho_A^n)$ with $\t\rho_A = \Tr_{\cal B} \ket{\Psi_R}\bra{\Psi_L}$, as these two eigenstates are generically different. However, with ${\tt PT}$ symmetry, and because of chiral factorization, we expect the equality $\ket{\Psi_R} = \ket{\Psi_L}$ to hold at quantum critical points, in many CFT models. Indeed, consider the field $\phi$ associated to the ground state. By chiral factorization it is a product of chiral and anti-chiral fields, $\varphi\b\varphi$ (this is true in all minimal models, but in general it may be a linear combination of such terms). Within radial quantization centered on the field, with coordinate $z=e^{ix+\tau}$, the transformation $z\mapsto \b z$ is a parity transformation. This transformation exchanges $\varphi\leftrightarrow\b\varphi$, which preserves $\phi$ (in general, the linear combination is required to be symmetric). Hence, the ground state is parity invariant. By ${\tt PT}$ symmetry, it is ${\tt T}$-invariant, and since the ${\tt T}$ transformation in general maps left and right eigenvectors to each other, this shows the claim. Certainly a more precise analysis would be interesting, but we have numerically verified this for the model (\ref{ham}).

\begin{figure}[h]
 \begin{center} \includegraphics[width=6cm]{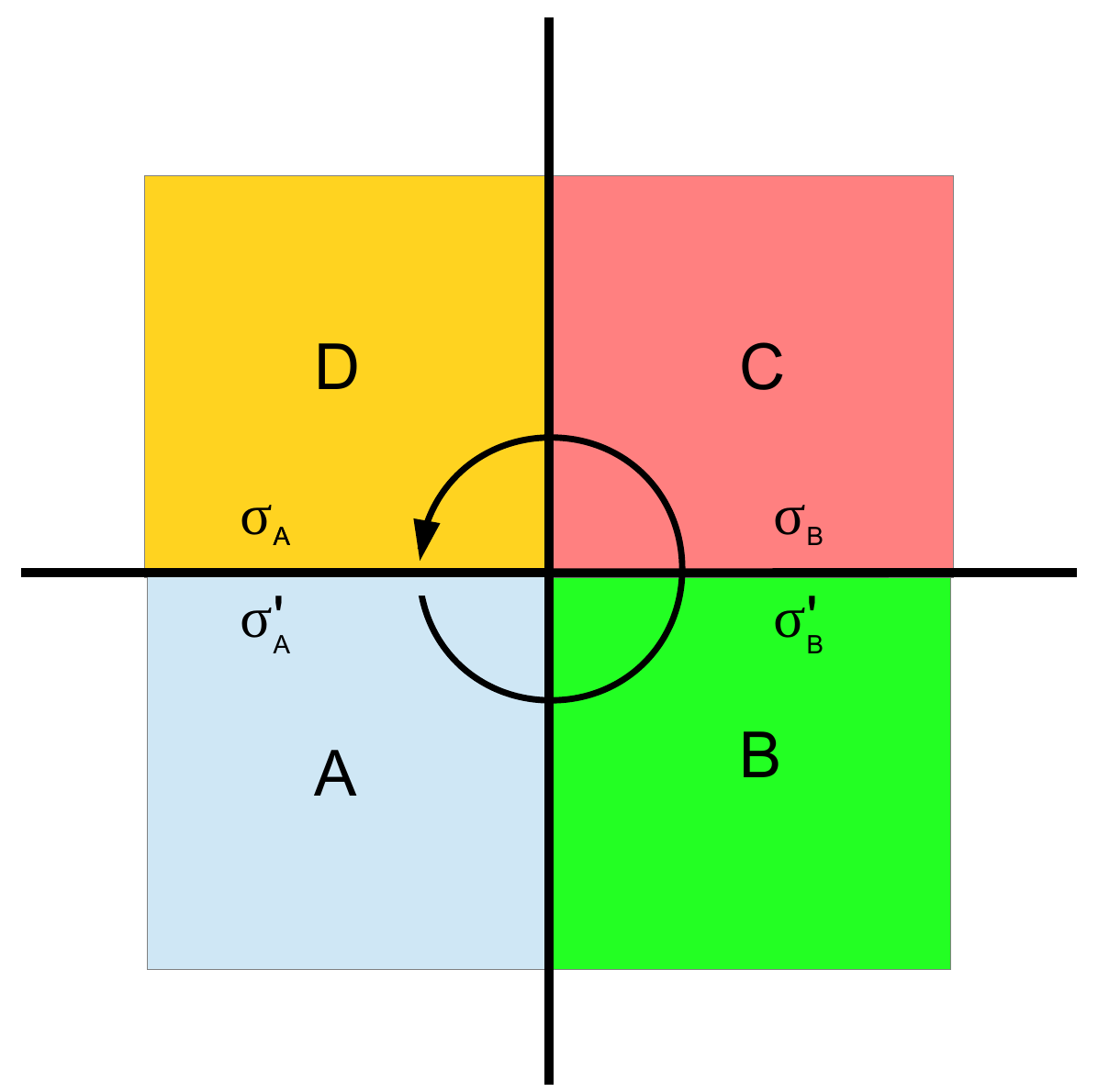} 
 \end{center} 
 \caption{Let the horizontal axis be a one-dimensional quantum system, then multiplication of the four CTMs gives the reduced density matrix.} 
 \label{fig:corner} 
 \end{figure}

\section{Model analysis}

Consider an off-critical infinite system, with a subsystem $A$ from the origin to infinity. For integer $n$ the quantity $\Tr_{\cal A}(\rho_A^n)$ may be evaluated using the corner transfer matrix (CTM) approach \cite{Peschel} within the associated statistical model on a lattice.
The CTMs $A$, $B$, $C$ and $D$ were introduced by Baxter \cite{BaxterBook}: they are the partition functions of the four corners of the lattice with fixed states on the edges of the corners (see FIG.~\ref{fig:corner}).

The reduced density matrix is proportional to the product $ABCD$  \cite{Peschel}. For integrable models, the eigenvalues are known using Yang-Baxter equations, and one may evaluate $\Tr_{\cal A}(\rho_A^n)$ explicitly. Generalizing previous results \cite{Franchini}, we have evaluated the R\'enyi entropy (\ref{def2}) for the Forrester-Baxter (FB) Restricted Solid On Solid Model (RSOS) on the square lattice \cite{ForresterBaxter}. This model is a lattice realisation of all off-critical minimal models $\mathcal{M}_{p,p'}(\xi)$, including the non-unitary series. Having real (but not necessarily positive) Boltzmann weights, it satisfies the requirement that $\ket{\Psi_R} = \ket{\Psi_L}$. We recovered (\ref{nus}) with $\ell/\ep$ replaced by the correlation length $\xi$ \cite{extension}.

\begin{figure}[h]
\begin{center}
\includegraphics[width=0.5 \columnwidth]{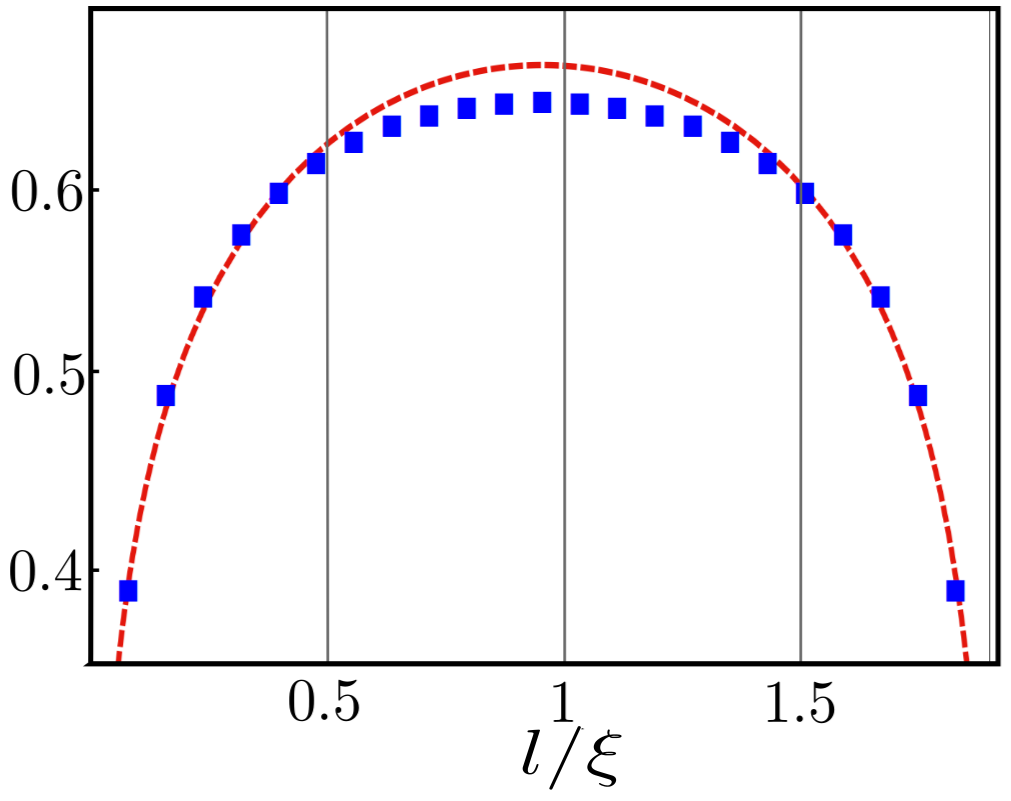}
\end{center}
\caption{Entanglement entropy for $\lambda=0.9$ and $L=24$. The dashed  line is the fitting curve $\frc{c_{\rm{eff}}}3\log\left(\frac{L}{\pi}\sin\lt(\frac{\ell \pi} {L}\rt)\right)+ \alpha$ with $c_{\mathrm{eff}}=0.4056$, and $\alpha=0.3952$. The agreement is remarkable for $\ell<\xi/2$.}
       \label{figure3}
       \end{figure}

We have also numerically evaluated the von Neumann EE of the ground state of (\ref{ham}). This is associated to the Lee-Yang minimal CFT, with $c_{\rm eff} = 2/5$. We exactly diagonalized (\ref{ham}) at various critical points. The critical line is defined by the values $(\lambda,h)=(\lambda,h_c(\lambda))$ such that, at $L=\infty$, the ground state energy $E_\phi(L)$ and the first excited state energy $E_0(L)$ coincide. The model possesses a non-trivial velocity of sound $v(\lambda,h)$, and at finite $L$ one expects the critical behaviour for $\ell\ll \xi(L)=v(\lambda,h_c(\lambda))/\left(E_0(L)-E_\phi(L)\right)$.
A thorough study of the entropy for $\lambda=0.9$ and chains of length up to $L=24$ gives the results reported in FIG.~\ref{figure3}. We confirm with great accuracy that (\ref{cftL}) (with $n=1$) holds with $c \mapsto c_{\rm{eff}}$. 

\section{Conclusion}

In this paper we have provided an analytic derivation using CFT techniques as well as numerical confirmation of the scaling (\ref{nus}), its logarithmic extension \eqref{reslog}, and its generalization beyond criticality and for finite systems. The main conclusions are that for quantum systems which possess a critical point described by a non-unitary conformal field theory, the logarithmic scaling of the entanglement entropy is characterized by the effective central charge $c_{\text{eff}}$ rather than the central charge $c$ (which for may non-unitary theories would be negative or zero) and that additional $\log(\log(\ell))$ corrections may arise in logarithmic CFTs. 

Although the result seems rather natural and reminiscent of the behaviour of the ground state energy of non-unitary critical systems, it is a powerful result in the context of entanglement. The entanglement entropy provides universal information about the nature of critical points, but that information (in general) is the effective central charge, rather than the central charge. Hence, for many non-unitary systems, such as the non-unitary minimal models, the entanglement entropy is encapsulating information both about the central charge and about the lowest dimensional field in the theory. As far as we are aware, there is still no physical quantity that extracts the central charge itself in non-unitary CFT models.

Besides the observation above it is also important to emphasize how the results have been obtained in the context of CFT. On the analytical side, we have used an approach which stands half-way between the methods of Holzhey, Larsen and Wilczek \cite{Holzhey} and those of Calabrese and Cardy \cite{CalabreseCardy}. Our approach is more algebraic in that it relies on  the generators (\ref{LL}) which provide a realization of the Virasoro algebra on the $\mathbb{Z}_n$ orbifold CFT. The use of this realization leads to the occurrence of $c_{\text{eff}}$ in a very natural way, as well as to the extra double-logarithmic term for logarithimic representations (wilst also leading to the usual results in the unitary case). They also incorporate in a natural way the dimension of the twist field whose correlators generate partition functions on the orbifold theory. Interestingly the dimension that occurs naturally in our computation is that of the lowest dimensional twist field that can be constructed in the theory. This has allowed us to also generalize the twist field description of the entanglement entropy \cite{CalabreseCardy, entropy} to non-unitary theories, and we believe this may form the basis for a deeper understanding of the CFT of non-unitary models.

There are various interesting problems that we would like to pursue in the future. Certainly, it is natural to consider the generalization of our results to quantum systems beyond criticality, such as massive integrable models \cite{inprogress}. Another idea is that the fact that the natural object to describe entanglement in non-unitary theories seems to be a ratio of correlation functions is very suggestive; it is interesting to ask whether this may provide a more general prescription on how to construct physical correlators in non-unitary models. We are currently pursuing these paths.

\medskip

\noindent {\bf{Acknowledgments}} D.B. thanks E.~Ercolessi for discussions and City University London for his doctoral studentship. B.D. thanks Universit\'e Paris Diderot, where part of this work was done, for financial support through a visiting professorship. E.L. thanks M.~Marcuzzi, J.~Hickey and A.~James for insightful discussions. F.R. thanks the I.N.F.N. grant GAST for partial financial support and F.~Franchini, E.~Ercolessi and D.~Palazzo for useful discussions. O.C.-A., B.D. and F.R. thank the organizers of the workshop on Quantum Integrability, Conformal Field Theory and Topological Quantum Computation which took place in Natal (Brazil) in March of 2014, where important discussions relating to this project took place. We thank J. Bhaseen and A.~Fring for their valuable comments.


\begin{thebibliography}{0}

\bibitem{white} S.R.~White, Phys. Rev. Lett. {\bf 69}, 2863 (1992).

\bibitem{vidal} G.~Vidal, Phys. Rev. Lett. {\bf 101}, 110501 (2008).

\bibitem{bennet}
C.~H. Bennett, H.J.~Bernstein, S.~Popescu, and B.~Schumacher,
\newblock Phys. Rev. {\bf A53}, 2046 (1996).

\bibitem{catalysis} D.~Jonathan and M.B.~Plenio, Phys. Rev. Lett.
\textbf{83}, 3566 (1999).


\bibitem{majorization}  M.A.~Nielsen, Phys. Rev. Lett. {\bf 83}, 436 (1999); M.A.~Nielsen; S. Turgut, J. Phys. {\bf A40} 12185 (2007); M.~Klimesh, Proc. 2004 International Symposium on Information Theory (ISIT 2004), June 27--July 2, 2004, p. 357.

\bibitem{lehur} K.~Le Hur, Ann. Phys. {\bf 323}, 2208 (2008).

\bibitem{hep} L.~Bombelli, R.K.~Koul, J.~Lee, and R.D.~Sorkin, Phys. Rev. {\bf D34}, 373 (1986). 

\bibitem{review} P.~Calabrese, J.L.~Cardy and B.~Doyon (ed), J. Phys. {\bf A42} 500301 (2009).


\bibitem{qsc} J.P.~Goff, D.A.~Tennant, and S.E.~Nagler,
Phys. Rev. \textbf{B52}, 15992 (1995); K.~Totsuka,
Phys. Rev. \textbf{B57}, 3454 (1998); B.C.~Watson, V.N.~Kotov, M.W.~Meisel, D.W.~Hall, G.E.~Granroth, W.T.~Montfrooij, S.E.~Nagler, D.A.~Jensen, R.~Backov, M.A.~Petruska, G.E.~Fanucci and D.R.~Talham,
Phys. Rev. Lett. \textbf{86}, 5168 (2001); B.~Thielemann, Ch.~R\"uegg, H.M.~R{\o}nnow, A.M.~L\"auchli, J.-S.~Caux, B.~Normand, D.~Biner, K.W.~Kr\"amer, H.-U.~G\"udel, J.~Stahn, K.~Habicht, K.~Kiefer, M.~Boehm, D.F.~McMorrow and J.~Mesot,
Phys. Rev. Lett. \textbf{102}, 107204 (2009); A.B.~Kuklov and B.V.~Svistunov,
Phys. Rev. Lett. \textbf{90}, 100401 (2003); L.-M.~Duan, E.~Demler, and M.D.~Lukin,
Phys. Rev. Lett. \textbf{91}, 090402 (2003); J.J.~Garc\'ia-Ripoll and J.I.~Cirac,
Phys.\textbf{5}, 76 (2003); M.~Lewenstein,  A.~Sanpera, V.~Ahufinger, B.~Damski, A.~Sen(De) and U.~Send,
Adv. Phys. \textbf{56}, 243 (2007); Y.-A.~Chen, S.~Nascimb\`ene, M.~Aidelsburger, M.~Atala, S.~Trotzky and I.~Bloch, Phys. Rev. Lett. \textbf{107}, 210405 (2011). T.~Kinoshita, T.~Wenger and D.~S.~Weiss, Nature \textbf{440}, 900 (2006).

\bibitem{Holzhey} C.~Holzhey, F.~Larsen and F.~Wilczek, Nucl. Phys. {\bf B424}, 443467 (1994).

\bibitem{Latorre1}
G.~Vidal, J.I.~Latorre, E.~Rico, and A.~Kitaev,
\newblock Phys. Rev. Lett. {\bf 90}, 227902 (2003).

\bibitem{Latorre2}
J.I.~Latorre, E.~Rico, and G.~Vidal,
\newblock Quant. Inf. Comput. {\bf 4}, 48 (2004).

\bibitem{CalabreseCardy}
P.~Calabrese and J.~L.~Cardy, J. Stat. Mech. (2004) P06002; J. Stat. Mech. (2005) P04010.

\bibitem{cthzam}  A.B.~Zamolodchikov, JETP Lett {\bf 43}, 730 (1986).

\bibitem{arealaw} L.~Bombelli, R.K.~Koul, J.~Lee, and R.D.~Sorkin, Phys.Rev. {\bf D34}, 373 (1986). J.~Eisert, M.~Crammer and M.B.~Plenio, Rev. Mod. Phys. {\bf 82}, 277 (2010).

\bibitem{german} F.~Castilho Alcaraz, M.~Ib\'a\~nez Berganza, G.~Sierra, Phys. Rev. Lett. \textbf{106}, 201601 (2011); M.~Ib\'a\~nez Berganza, F.~Castilho Alcaraz and G.~Sierra, J. Stat. Mech. (2012) P01016. 

\bibitem{fab} F.H.L.~Essler, A.M.~L\"auchli, P.~Calabrese
Phys. Rev. Lett. {\bf 110}, 115701 (2013). P.~Calabrese, 
F.H.L.~Essler and A.M.~L\"auchli, J. Stat. Mech. (2014) P09025. 

\bibitem{Knizhnik} V.G~Knizhnik, Commun. Math. Phys. \textbf{112}, 567 (1987).

\bibitem{entropy}
J.~L. Cardy, O.~A. Castro-Alvaredo and B.~Doyon,
\newblock J. Stat. Phys. {\bf 130}, 129 (2008).

\bibitem{permutation} O.A.~Castro-Alvaredo and B.~Doyon, J. Stat. Mech. (2011) P02001. 

\bibitem{negativity} P.~Calabrese, J.L.~Cardy and E.~Toni, Phys. Rev. Lett. {\bf 109}, 130502 (2012); J. Stat. Mech. (2013) P02008; P.~Calabrese, L.~Tagliacozzo and E.~Toni,  J. Stat. Mech. (2013) P05002. 

\bibitem{cft} P.~Di Francesco, P.~Mathieu and D.~S\'en\'echal, Conformal Field Theory, Springer (1997). A.A.~Belavin, A.M.~Polyakov and A.B.~Zamolodchikov, Nucl. Phys. {\textbf{B241}} 333 (1984).

\bibitem{vjs} E.~Vernier, J.L.~Jacobsen and H.~Saleur, J. Phys. {\bf A47}, 285202 (2014).

\bibitem{ef} K. Efetov, Supersymmetry in Disorder and Chaos, Cambridge University Press, Cambridge, UK, 1999

\bibitem{gu} V. Gurarie, Nucl. Phys. {\bf B410}, 535 (1993); Nucl. Phys. {\bf B546}, 765 (1999).

\bibitem{ms} Z. Maassarani and D. Serban, Nucl. Phys. {\bf B489}, 603-625 (1997).

\bibitem{ali} M.~Alishahiha, A.F.~Astaneh, and M.R.M.~Mozaffar, Phys. Rev. {\bf D89}, 065023 (2014).


\bibitem{PS} F.C.~Alcaraz, M.N.~Barber, M.T.~Batchelor, R.J.~Baxter, and G.R.W.~Quispel, J. Phys. \textbf{A20}, 6397 (1987);
V.~Pasquier and H.~Saleur, Nucl. Phys. \textbf{B330}, 523 (1990);
G.~J\"uttner and M. Karowski, Nucl. Phys. \textbf{B430}, 615 (1994). C.~Korff and R.A.~Weston, J.Phys. {\bf A40}, 8845 (2007).


\bibitem{vgehlen} G.~von Gehlen,
J. Phys. \textbf{A24} 5371 (1991); Int. J. Mod. Phys. \textbf{B8} 3507 (1994).


\bibitem{yl1} M.E.~Fisher, Phys. Rev. Lett. {\bf 40}, 1610 (1978);
J.L.~Cardy, Phys. Rev. Lett. {\bf 54}, 1354 (1985).

\bibitem{bender}
C.M.~Bender, Rept. Prog. Phys. {\bf 70}, 947 (2007); C.~Figueira de Morisson Faria and A.~Fring, Laser Physics, {\bf 17}, 424 (2007); A. Mostafazadeh, Int. J. Geom. Meth. Mod. Phys. {\bf 7}, 119 (2010). 

\bibitem{sgh} F.G. Scholtz, H.B. Geyer, F.J.W. Hahne, Ann. of Phys. {\bf 213}, 74 (1992).

\bibitem{pt} C.M.~Bender and S.~Boettcher, Phys. Rev. Lett.
{\bf 80}, 5243 (1998).

\bibitem{revpt} A.~Fring, Phil. Trans. R. Soc. A {\bf 371}, 20120046 (2013).


\bibitem{meandreas} O.A.~Castro-Alvaredo and A.~Fring, J. Phys. \textbf{A42} 465211 (2009).

\bibitem{wow} 
C.E.~R\"uter, K.G.~Makris, R.~El-Ganainy, D.N.~Christodoulides, M.~Segev and D.~Kip, Nature Phys. {\bf 6}, 192 (2010); L.~Feng, M.~Ayache, J.~Huang, Y.-L.~Xu, M.-H.~Lu, Y.-F.~Chen, Y.~Fainman, A.~Scherer, Science {\bf 333}, 729 (2011); A~Regensburger,	C.~Bersch,M.-A.~Miri,G.~Onishchukov,D.N.~Christodoulides and U.~Peschel, Nature {\bf 488}, 167 (2012). 


\bibitem{exper} T.~Eichelkraut, R.~Heilmann, S.~Weimann, S.~Stützer, 
F.~ Dreisow, D.N.~Christodoulides, S.~Nolte	
and A.~Szameit, Nature Commun. {\bf 4}, 2533 (2013).

\bibitem{nott} J.M.~Hickey, S.~Genway, I.~Lesanovsky and J.P.~Garrahan, Phys. Rev. {\bf B87}, 184303 (2013). 


\bibitem{timeint}
J.M.~Hickey, E.~Levi and J.P.~Garrahan, Phys.Rev. {\bf B90}, 094301 (2014).


\bibitem{lon} S.~Longhi, Phys. Rev. Lett. {\bf 105}, 013903 (2010).

\bibitem{bw} U. Bilstein and B. Wehefritz, J. Phys. {\bf A30}, 4925 (1997).


\bibitem{huck} B. Huckenstein, Rev. Mod. Phys. {\bf 67}, 357 (1995).

\bibitem{nest} A.I.~Nesterov, G.P.~Berman, J.C.~ Beas Zepeda and A.R.~Bishop, Quant. Inf. Process. {\bf 13}, 371 (2014).


\bibitem{ceff} C.~Itzykson, H.~Saleur, and J.-B.~Zuber, Europhys. Lett. \textbf{2}, 91 (1986).


\bibitem{affleck} I.~Affleck, Phys. Rev. Lett. \textbf{56} 746--748 (1986).

\bibitem{BCN} H.W.J. Blote, J.L.~Cardy and M.P.~Nightingale, Phys. Rev. Lett. \textbf{56} 742--745 (1986).


\bibitem{ctheorem} O.A.~Castro-Alvaredo, B.~Doyon and E.~Levi,
J. Phys. \textbf{A44} 492003 (2011); E.~Levi, J. Phys. \textbf{A45} 275401 (2012).

\bibitem{DDH} B.~Doyon, M.~Hoogeveen and D.~Bernard, J. Stat. Mech. (2014) P03002. 


\bibitem{extension} D.~Bianchini, Entanglement entropy in restricted integrable spin chains, MSc Thesis, University of Bologna (2013); D.~Bianchini and F.~Ravanini, in preparation.


\bibitem{ForresterBaxter} P.~J.~Forrester and R.~J.~Baxter, J.
Stat. Phys. \textbf{38}, 435--472 (1985).
 

\bibitem{Peschel} L.~ Kaulke, I.~Peschel and M.~Legeza, Annalen der Physik 8(2), 153--164 (1999); I.~Peschel, J. Stat. Mech. (2004) P12005.


\bibitem{BaxterBook} R.J.~Baxter, Exactly solved models in statistical mechanics, London Academic Press Inc. (1982).

\bibitem{Franchini} A.~De Luca and F.~Franchini, Phys. Rev. \textbf{B87}, 045118 (2013).

\bibitem{KacWakimoto} V.~G.~Kac and M.~Wakimoto, Acta Appl. Math. \textbf{21} 3, (1990) 


\bibitem{Bouwknegt} P.~Bouwknegt, Coset construction for winding subalgebras and applications, q-alg/9610013 (1996).


\bibitem{Borisov} L.~Borisov, M.~B.~Halpern and C.~Schweigert, Int. J. Mod. Phys. \textbf{A13}, 125-168 (1998). 



\bibitem{inprogress} D.~Bianchini, O.~Castro-Alvaredo and B.~Doyon, work in progress.



\end{thebibliography}
\end{document}